\documentclass[pre,preprint,aps,eqsecnum]{revtex4}

\usepackage{psfrag}
\usepackage{graphicx}
	\psfrag{stress_s}{\Huge Stress $s$}	
     	\psfrag{strain}{\Huge Strain $\epsilon^{total}/
	\epsilon_{0}$}
	\psfrag{energy_ps}{\Huge Energy $\psi$}
	\psfrag{dl}{\Huge $\Delta$}	
        \psfrag{timeL}{\Huge Time}	
\begin{document}

\title{Dynamics of Shear-Transformation Zones in Amorphous Plasticity: Energetic Constraints in a Minimal Theory}

\author{J. S. Langer}
\author{L. Pechenik}

\affiliation{Department of Physics,
University of California,
Santa Barbara, CA  93106-9530  USA}
\date{July, 2003}

\begin{abstract}
We use energetic considerations to deduce the form of a previously uncertain coupling term in the shear-transformation-zone (STZ) theory of plastic deformation in amorphous solids.  As in the earlier versions of the STZ theory, the onset of steady deformation at a yield stress appears here as an exchange of dynamic stability between jammed and plastically deforming states.  We show how an especially simple ``quasilinear'' version of this theory accounts qualitatively for many features of plasticity such as yielding, strain softening, and strain recovery. We also show that this minimal version of the theory fails to describe certain other phenomena, and argue that these limitations indicate needs for additional internal degrees of freedom beyond those included here. 
\end{abstract}
\maketitle

\section{Introduction}

In developing a ``shear-transformation-zone'' (STZ) model of plasticity in noncrystalline solids, we have encountered several fundamental questions that pertain to the way in which mechanical work done on the system is stored reversibly and dissipated irreversibly during plastic deformation.  We find that the constraints imposed on our phenomenological theory by such considerations, plus one simple assumption, resolve an earlier uncertainty about the STZ theory and, in fact, determine essentially all the details of its simplest version.  With this assurance about the theory's internal self-consistency, we can look more carefully at the observed phenomena to determine what additional physical ingredients are needed to achieve quantitative predictive capabilities within this basic framework.  The present paper provides an account of the first stages of that investigation.

A few preliminary comments may be useful. We recognize that the conventional approaches to plasticity theory have, for almost a century, been extremely successful in engineering applications. There are, however, some puzzling internal inconsistencies that pervade all of solid mechanics and that will have to be resolved if this field is to meet modern technological challenges. Questions of this general nature seem certain to arise in attempts to understand other strongly nonequilibrium phenomena such as those that occur in geology, polymer science, and especially biology.  

The most basic of these questions is: What are the fundamental distinctions between brittle and ductile behaviors? A brittle solid breaks when subjected to a large enough stress, whereas a ductile material deforms plastically.  Remarkably, we do not yet have a fundamental understanding of the distinction between these two behaviors. Conventional theories of crystalline solids say that dislocations form and move more easily through ductile materials than brittle ones, thus allowing deformation to occur in one case and fracture in the other.  But the same behaviors also occur in amorphous solids; thus the dislocation mechanism cannot be the essential ingredient of all theories.  Moreover, the brittleness or ductility of some materials depends upon the speed of loading, which implies that a proper description of deformation and fracture must be dynamic, that is, it must be expressed in the form of equations of motion rather than the conventional phenomenological rules and yielding criteria.\cite{LUBLINER} 

A second fundamental question is: What is the origin of memory effects in plasticity? Standard, hysteretic, stress-strain curves for deformable solids tell us that these materials have rudimentary memories.  Roughly speaking, they ``remember'' the direction in which they most recently have been deformed. When unloaded and then reloaded in the original direction, they are hardened and respond only elastically, whereas, when loaded in the opposite direction, they deform plastically.  The conventional way of dealing with such behavior is to specify phenomenological rules stating how the response to an applied stress is determined by the history of prior loading; but such rules provide little insight about what is actually happening or what might be the nature of a theory based more directly on molecular mechanisms.  

A better way to deal with memory effects is to introduce internal state variables that carry information about previous history and determine the current response of the system to applied forces.  All too often, however, the plastic strain itself is used as such a state variable -- a procedure that violates basic principles of nonequilibrium physics because it implies that a material must somehow remember {\it all} of its prior history of deformation starting from some primordial reference state. That cannot be possible for an amorphous solid any more than it is for a liquid, where it is well understood that only displacement rates, and not the displacements themselves, may appear in equations of motion.  Nevertheless, the preference for Lagrangian formulations that specify plastic displacements relative to a permanently fixed reference state permeates a large part (but not all \cite{RICE}) of the literature on plasticity.  We strongly prefer to develop Eulerian formulations with appropriate internal state variables, as we have done in \cite{ELP}. The question remains, then:  What are the appropriate variables for amorphous solids? 

The STZ theory that we shall discuss here is an attempt to identify those state variables and their equations of motion. The original ideas are largely due to M. Falk \cite{FL,FLMRS,Fthesis}, who used molecular dynamics simulations of shear deformations in two-dimensional, amorphous, Lennard-Jones solids to show that, as postulated by Cohen, Turnbull, Spaepen, Argon and others \cite{TURNBULL,SPAEPEN}, irreversible deformations are localized in dilute distributions of ``shear-transformation zones.''  Falk showed that these zones behave like two-state systems.  That is, in the presence of a shear stress, they can deform by only a finite amount in one direction before they become jammed but, once they have done so, they can transform in the opposite direction in response to a reversed stress.  The STZ's are ephemeral; they are created and annihilated during irreversible deformations of the material. This picture implies that the relevant state variables are the population densities of the STZ's in their various orientations. The equations of motion for these populations have interesting implications, the most important of which is the notion that the onset of steady deformation at a yield stress occurs as an exchange of dynamic stability between jammed (non-deforming) and unjammed (deforming) states of the system.  Section II of this paper contains a brief review of the original ideas and the way in which they are specialized for use here in a minimal but useful version of an STZ theory of amorphous plasticity.

In Section III, we show how the constraints imposed by the first and second laws of thermodynamics determine the structure of the equations of motion for the STZ state variables.  We argue that the rate of energy dissipation during deformation must be proportional to the rates at which STZ's are annihilated and created.  With this hypothesis, we compute both the dissipation rate and the recoverable energy stored in the plastic degrees of freedom.  

Finally, in Section IV, we discuss some implications of these results. We compute theoretical stress-strain curves for systems driven both at constant strain rates and at fixed stresses (creep tests).  Our goal here is to demonstrate qualitatively the wide range of phenomena that are described by this theory, and also to show what qualitative features are missing.  We conclude by making some remarks about the basic ingredients of a more complete dynamical theory of amorphous plasticity.  

\section{Summary of STZ Dynamics}

As in \cite{FL}, we consider only strictly two-dimensional non-crystalline systems. We further restrict ourselves to molecular materials in contact with  thermal reservoirs, so that we may assume that an ambient temperature determines an underlying fluctuation rate which, in turn, determines the rates at which the molecules explore their configurations.  Thus, we shall not (for the present) consider granular materials or foams where ordinary thermal kinetic energies are negligibly small, and where the motions of the particles during rearrangements must be driven entirely by the external forces applied to the system. 

We consider here only situations in which the orientation of the  stress and strain tensors remains fixed.  A tensorial version of this theory, applicable to more general situations where the stresses rotate during plastic deformation, has been used in our earlier studies of microstructural shear banding \cite{SHEARLOC} and necking instabilities \cite{ELP}, and is developed in more detail in \cite{PECHENIK}.  With the restriction of fixed stress orientation, it is sufficient to assume that the population of STZ's consists simply of zones oriented along the principal axes of the two-dimensional stress tensor.  It is shown in \cite{PECHENIK} that exactly the same equations as the ones we shall use here can be derived starting from the assumption that the {\it a priori} orientations of the zones is circularly symmetric.

Let the deviatoric stress be diagonal along the $x$, $y$ axes; specifically, let $s_{xx}=-s_{yy}= s$ and $s_{xy}=0$. Then choose the ``$+$" zones to be oriented (elongated) along the $x$ axis, and the ``$-$" zones along the $y$ axis; and denote the population density of zones oriented in the ``$+$''/``$-$'' directions by the symbol $n_{\pm}$.  With these conventions, the plastic strain rate is:
\begin{equation}
\label{epsdot}
\dot\epsilon_{xx}^{pl}= -\dot\epsilon_{yy}^{pl}\equiv \dot\epsilon^{pl}=\lambda\,\Bigl(R_-(s)\,n_--R_+(s)\,n_+\Bigr).
\end{equation}
Here $\lambda$ is a material-specific parameter with the dimensions of $({\rm length})^2$, which must be roughly equal to the area of an STZ, that is, a few  square molecular spacings.  The quantity in parentheses in Eq.(\ref{epsdot}) is the net rate per unit area at which STZ's are transforming from ``$-$'' to ``$+$'' orientations.  Here, $R_+(s)$ and $R_-(s)$ are the rates for ``$+$" to ``$-$'' and ``$-$" to ``$+$'' transitions respectively. For simplicity, we write these rates as explicit functions of only the deviatoric stress $s$, although they depend implicitly on the temperature and pressure and perhaps other quantities.  

The equations of motion for the populations $n_{\pm}$ must have the form:
\begin{equation}
\label{ndot}
\dot n_{\pm}=R_{\mp}(s)\,n_{\mp}-R_{\pm}(s)\,\,n_{\pm}  +\Gamma(s,...)\,\left({n_{\infty}\over 2}-n_{\pm}\right),
\end{equation}
where the last two terms in parentheses, proportional to $\Gamma$, describe creation and annihilation of STZ's.  Here, $n_{\infty}$ is the total density of zones that would be generated in a system that is undergoing steady plastic deformation.  Introducing $n_{\infty}$ in Eq.(\ref{ndot}) is simply a way to characterize the ratio of the creation and annihilation rates in terms of a physically meaningful quantity.  The factor $\Gamma$ that determines these rates is a function of the stress and the strain rate or, equivalently, the stress and the population densities.  The choice of $\Gamma$ is one of the principal topics of this paper; it is discussed in detail in Section III.      

We define dimensionless internal state variables by writing
\begin{equation}
\label{vardef}
\Lambda \equiv {n_++n_-\over n_{\infty}},~~~~\Delta\equiv {n_+-n_-\over n_{\infty}}.
\end{equation}
These quantities, $\Lambda$ and $\Delta$, are the internal state variables, or order parameters, that we believe are appropriate for a dynamical theory of amorphous plasticity.  In a more general treatment \cite{PECHENIK}, $\Lambda$ remains a scalar density, but $\Delta$ becomes a traceless symmetric tensor with the same transformation properties as the deviatoric stress.  We also define:
\begin{equation}
\label{Tdef}
{\cal S}\equiv {1\over 2}\,(R_--R_+),~~~~{\cal C}\equiv {1\over 2}\,(R_-+R_+),~~~{\cal T}\equiv {{\cal S}\over{\cal C}}.
\end{equation}
Then the STZ equations of motion become:
\begin{equation}
\label{doteps}
\dot\epsilon^{pl}=\epsilon_0\,{\cal C}(s)\,\Bigl(\Lambda\,{\cal T}(s)-\Delta\Bigr);
\end{equation}
\begin{equation}
\label{dotdelta}
\dot\Delta=2\,{\cal C}(s)\,\Bigl(\Lambda\,{\cal T}(s)-\Delta\Bigr)-\Gamma(s,\Lambda,\Delta)\,\Delta;
\end{equation}
and 
\begin{equation}
\label{dotlambda}
\dot\Lambda=\Gamma(s,\Lambda,\Delta)\,\Bigl(1-\Lambda \Bigr).
\end{equation}
Here, we have defined $\epsilon_0 \equiv\lambda\,n_{\infty}$.  This is the only material-specific parameter remaining explicitly in these equations.  $\epsilon_0$ is roughly the fraction of the total area of the system covered by the STZ's; therefore, to be consistent with our basic assumptions, it must be much smaller than unity.

Throughout the rest of this paper, we shall use only what we call the ``quasilinear'' version of these equations.\cite{FLMRS}  That is, we write:
\begin{equation}
\label{quasilinear}
{\cal T}(s)\cong s;~~~~{\cal C}(s)\cong 1;
\end{equation}
so that Eqs.(\ref{doteps}) and (\ref{dotdelta}) become:
\begin{equation}
\label{dotepsql}
\dot\epsilon^{pl}=\epsilon_0\,(\Lambda\,s-\Delta);
\end{equation}
\begin{equation}
\label{dotdeltaql}
\dot\Delta=2\,(\Lambda\,s-\Delta)-\Gamma(s,\Lambda,\Delta)\,\Delta.
\end{equation}
We have written the right-hand sides of Eqs.(\ref{quasilinear}) without factors whose dimensions would be, respectively, inverse stress and inverse time.  This means that, without loss of generality, we are implicitly expressing all stresses and (later) elastic moduli in units of some unspecified characteristic stress. That characteristic stress will turn out to be the dynamic yield stress, which implicitly contains the temperature and pressure dependence of the rates $R_{\pm}$. Similarly, we have set the unit of time equal to the inverse of the rate factor contained in the function ${\cal C}(s)$.    

Note that the quasilinear version of the STZ theory looks directly comparable to some conventional phenomenology.\cite{LUBLINER}  In particular, the quantity $\Delta$ apparently plays the role of a ``back stress'' or a ``hardening parameter'' in Eq.(\ref{dotepsql}), although it has a different physical interpretation here than it does elsewhere.  If the nonlinear term $-\Gamma\,\Delta$ were missing on the right-hand side of Eq.(\ref{dotdeltaql}), then we would be able to integrate both sides of that equation over time and deduce that the ``back stress'' $\Delta$ is directly proportional to the total plastic strain.  The ephemeral nature of the STZ's, as expressed in $-\Gamma\,\Delta$, precludes any such interpretation except, perhaps, in situations where the plastic strain is so small that the nonlinear term is negligible.

The quasilinear theory has important advantages but also serious limitations.  On the negative side, when $s>1$, the linear approximation for ${\cal T}(s)$ in Eq.(\ref{quasilinear}) violates the inequality ${\cal T}(s)<1$ implied by the definitions in Eq.(\ref{Tdef}). This is a serious shortcoming if we are to take the STZ picture literally, that is, if we make the strong assumption that all the zones have the same size and interact with one another only in a mean-field sense.  If the latter conditions are not true, but if the basic picture of localized deformations remains valid, then the linear representation for ${\cal T}(s)$ might be qualitatively correct over a wider range of stresses, and the the quasilinear theory might have the merit of being the simplest description of dynamic plasticity consistent with the symmetries of the system and the choice of order parameters.  

Another limitation of the quasilinear theory is that it loses some of the STZ memory effects, specifically, those that reside in the stress dependence of ${\cal C}(s)$. This is an important topic that shall address in Section IV as part of a more general discussion of possible extensions of this theory.  

On the plus side, the quasilinear theory has the great advantage of simplicity. It is easy to interpret and to use in numerical calculations such as those reported in our recent study of the necking instability.\cite{ELP}  It may be the closest we can come to a description of deformable amorphous solids that is comparable in utility to the Navier-Stokes equations for fluid dynamics.  

\section{Energy Balance}

We turn now to the energetics of the quasilinear STZ model.  The introduction of the internal state variables $\Lambda$ and $\Delta$ raises the question of whether recoverable energy might be associated with these degrees of freedom and, if so, what the form of that energy function might be.  A related question is the relation between the state variables and the rate of energy dissipation during plastic deformation.  These are important questions; the energy stored in plastic degrees of freedom may, along with stored elastic energy, drive recovery of plastic strain.  That energy might also be partially recoverable, for example, during necking \cite{ELP} or fracture, thus affecting  estimates of failure rates or the Griffiths threshold. 

The energy-balance equation (the first law of thermodynamics) for this model has the form:
\begin{equation}
\label{energybalance}
2\,\dot\epsilon^{pl}\,s=
2\, \epsilon_0\,(\Lambda\,s-\Delta)\,s= \epsilon_0\,{d\over dt}\,\psi(\Lambda,\Delta) + {\cal Q}(s,\Lambda,\Delta).
\end{equation}
The left-hand side of Eq.(\ref{energybalance}) is the rate at which plastic work is being done.  On the right side, $\epsilon_0\,\psi$ is the state-dependent recoverable energy and ${\cal Q}$ is the dissipation rate.  ${\cal Q}$ must be positive in order for the system to satisfy the second law of thermodynamics, that is, for the work done in going around a closed cycle in state space to be positive. 

Next, consider the function $\Gamma(s,\Lambda,\Delta)$, which was defined in Eq.(\ref{ndot}) as determining the rates at which STZ's are annihilated and created.  In \cite{FL},  $\Gamma$ was chosen to be the rate at which plastic work is done on the system, that is, the left-hand side of Eq.(\ref{energybalance}).  As pointed out in \cite{FL}, this interpretation cannot be generally correct because the work rate can be negative (for example, during strain recovery), but the factor $\Gamma$ appearing in the creation and annihilation rates must always be positive or zero.  There are not many other simple choices for $\Gamma$, however.  On physical grounds, we expect $\Gamma$ to be quadratic in the driving force in a quasilinear theory such as this one.  Annihilation and creation of zones should be induced by local dilations or contractions, and dilational strain is a second-order response to shear stress.  The simplest non-negative possibility is $(\dot\epsilon^{pl})^2$, which has been explored in \cite{LL1}. As we shall see, the latter expression is close to being correct. 

On general grounds, we expect ${\cal Q}$ also to be quadratic in the driving force or, equivalently, in the strain rate; that is, we expect ${\cal Q}$ and $\Gamma$ to be similar functions.  We therefore propose that $\Gamma$ be the dissipation rate per STZ:
\begin{equation}
{\cal Q}(s,\Lambda,\Delta)= \epsilon_0\,\Lambda\,\Gamma(s,\Lambda,\Delta).
\end{equation}
With this hypothesis, we can use Eqs. (\ref{dotdeltaql}) and (\ref{dotlambda}) to write Eq.(\ref{energybalance}) in the form
\begin{equation}
2\,(\Lambda\,s-\Delta)\,s= {\partial\psi\over\partial\Lambda}\, \Gamma\,(1-\Lambda) + 2\,{\partial\psi\over\partial\Delta}\,(\Lambda\,s-\Delta-\Gamma\,\Delta)+\Lambda\,\Gamma.
\end{equation}
Then, solving for $\Gamma$, we find:
\begin{equation}
\label{Gamma1}
\Gamma={2\,(\Lambda\,s-\Delta)\,(s-\partial\psi/\partial\Delta)\over \Lambda+(1-\Lambda)(\partial\psi/\partial\Lambda)-\Delta\,(\partial\psi/\partial\Delta)}.
\end{equation}

We can assure positivity of the numerator in Eq.(\ref{Gamma1}) for all $s$ by choosing
\begin{equation}
{\partial\psi\over\partial\Delta}= {\Delta\over\Lambda},
\end{equation}
so that the numerator becomes $2\,\Lambda\,(s-\Delta/\Lambda)^2$.  Then, 
\begin{equation}
\psi(\Lambda,\Delta)={\Delta^2\over 2\,\Lambda}+\psi_0(\Lambda),
\end{equation}
where $\psi_0(\Lambda)$ is an as-yet undetermined constant of integration. We now have
\begin{equation}
\Gamma(s,\Lambda,\Delta)={2\,\Lambda\,(s-\Delta/\Lambda)^2\over M(\Lambda,\Delta)},
\end{equation}
where
\begin{equation}
M(\Lambda,\Delta)=\Lambda - (1+\Lambda)\,{\Delta^2\over 2\,\Lambda^2} + (1-\Lambda)\,{\partial\psi_0\over\partial\Lambda}.
\end{equation}

The second-law constraint requires that $M(\Lambda,\Delta)$ remain positive along all the system trajectories determined by our equations of motion in the space of variables $\Lambda$ and $\Delta$.  This happens automatically so long as all the trajectories start at points where $M(\Lambda,\Delta)>0$. The locus of points along which $M(\Lambda,\Delta)$ changes sign is a dynamical boundary for these trajectories; the dissipation rate diverges at that boundary, and the trajectories are strongly repelled from it in a way that does not allow them to cross into unphysical regions where the dissipation rate is negative. Our only free option, at this point, is to choose the function  $\psi_0(\Lambda)$.  If we let $\psi_0=\Lambda/2$, then 
\begin{equation}
M(\Lambda,\Delta)= {1\over 2}\,(1+\Lambda)\,(\Lambda^2-\Delta^2);
\end{equation}
\begin{equation}
\label{Gamma2}
\Gamma(s,\Lambda,\Delta)={4\,\Lambda\,(\Lambda\,s-\Delta)^2\over (1+\Lambda)\,(\Lambda^2-\Delta^2)};
\end{equation}
and
\begin{equation}
\psi(\Lambda,\Delta)={\Lambda\over 2}\,\left(1+{\Delta^2\over \Lambda^2}\right).
\end{equation}
Our special choice of $\psi_0(\Lambda)$ means that the inequality $\Delta^2 < \Lambda^2$, required by Eq.({\ref{vardef}), is saturated at the dynamical boundary.  Values of $\psi_0$ of the form $c\,\Lambda$ with $0< c \le 1/2$ remain consistent with the inequality and, so far as we can see, are not ruled out by our analysis.

To see what these results mean for the STZ dynamics, note first that the positivity of $\Gamma$ tells us that $\Lambda = 1$ is always the stable fixed point of Eq.(\ref{dotlambda}).  If we then let $\Lambda \to 1$ in Eq.(\ref{dotdeltaql}), we find
\begin{equation}
\label{dotdeltaql2}
\dot\Delta \to {2\,(s-\Delta)\,(1-s\,\Delta)\over 1-\Delta^2}.
\end{equation}
From this expression, it is clear by inspection that the jammed (non-deforming) steady state solution $s=\Delta$ is stable at fixed $s$ for $s<1$, and the unjammed (deforming) steady state solution $s=1/\Delta$ is stable for $s>1$.  This situation is illustrated in Fig. \ref{fig:flowdiagram}, 
\begin{figure}
      \includegraphics[angle=-90, width=0.7\columnwidth]{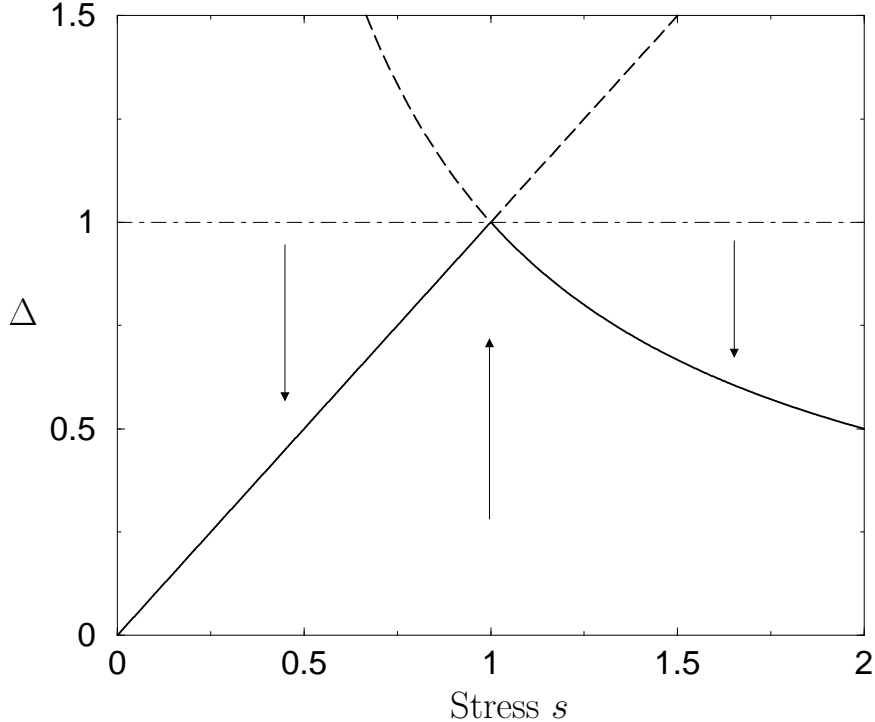}
      \caption{Locus of steady-state solutions of the STZ equations in the
      $s$-$\Delta$ plane.}
      \label{fig:flowdiagram}
\end{figure}
where the arrows in the figure indicate the sign of $\dot\Delta$ for fixed $s$.  The line $\Delta=1$ is the uncrossable boundary described above. This picture remains qualitatively correct in the more general situation where $\Lambda$ is allowed to vary, and even in circumstances where $s$ varies in response to controlled changes in the strain.  The exchange of stability between non-deforming and deforming states always occurs at $s=1$.

Our use of the term ``jamming''\cite{LIU-NAGEL} is intended to evoke a simple picture of the exchange of stability at the yield stress.  At small stresses, the system is literally jammed in the sense that the majority of the zones are oriented parallel to the applied stress and therefore are not able to contribute to further deformation in that direction.  The stable steady state is the one in which the strain rate and, accordingly, the rate of annihilation and creation of zones are all zero.  Above the yield stress, on the other hand, jammed zones are annihilated and new unjammed zones are created fast enough to sustain a stable, nonzero strain rate.  We shall examine these behaviors in more detail in the next Section. 

\section{Predictions and Limitations}

\begin{figure}
	\includegraphics[angle=-90, width=0.7\columnwidth]{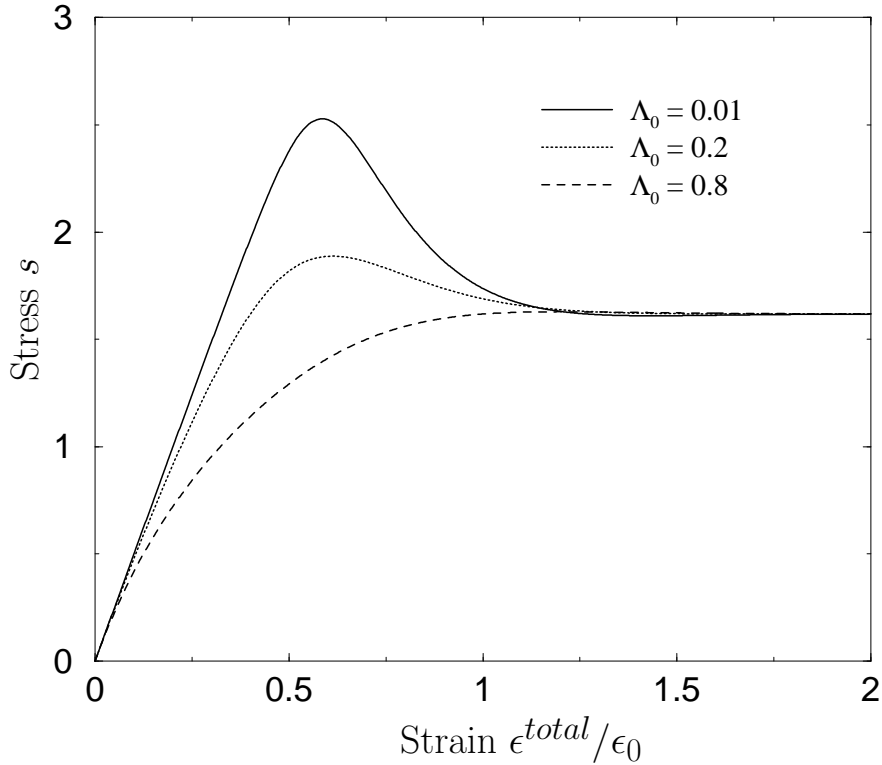}
\caption{Stress-strain curves for constant-strain-rate calculations for three
different initial densities of STZs.}
      \label{fig:stresses}
\end{figure}
To examine the predictions of this version of the STZ theory, we first consider simple experiments in which the stress is measured as the system undergoes pure shear at a constant total strain rate, say $\dot\epsilon^{total}= \epsilon_0\,q_0$. Here we make the same crucial simplifying assumptions that we have used in earlier work.  Specifically, we assume that the total strain rate $\dot\epsilon^{total}$, or more generally, the rate of deformation tensor, is the sum of elastic and plastic parts; and we further assume that the elastic rate of deformation is related to the rate of change of the stress by linear elasticity.  Thus, with the plastic strain rate given by Eq.(\ref{dotepsql}), the equation of motion for $s$ is
\begin{equation}
\label{q0}
\dot s = 2\,\mu\,\epsilon_0\,(q_0-\Lambda\,s+\Delta),
\end{equation}
where $\mu$ is the elastic shear modulus.  In Figs. 2 - 4, 
we show the results of solving Eq.(\ref{q0}) along with Eqs.(\ref{dotdeltaql}) and (\ref{dotlambda}) for $\Delta$ and $\Lambda$.  In all of these calculations, we have chosen $2\,\mu\,\epsilon_0=5$, which might correspond, for example, to $\epsilon_0\sim .025$ and $\mu \sim 100$.  Fig. 2 shows stress $s$ as a function of total strain for $q_0=1$ and for three different initial values of $\Lambda$, $\Lambda_0 = 0.01$, $0.2$ and $0.8$. The strain is shown in units of $\epsilon_0$, that is, in units roughly of order $10^{-2}$. In Fig. 3, we show the recoverable energy $\psi(\Lambda,\Delta)$, also as a function of total strain, for the same sets of parameters.  Fig. 4 illustrates the dependence on strain rate; that is, the three stress-strain curves shown there are for $q_0=.01$, $0.5$, and $1.0$, all for $\Lambda_0=0.5$.
\begin{figure}
      \includegraphics[angle=-90, width=0.7\columnwidth]{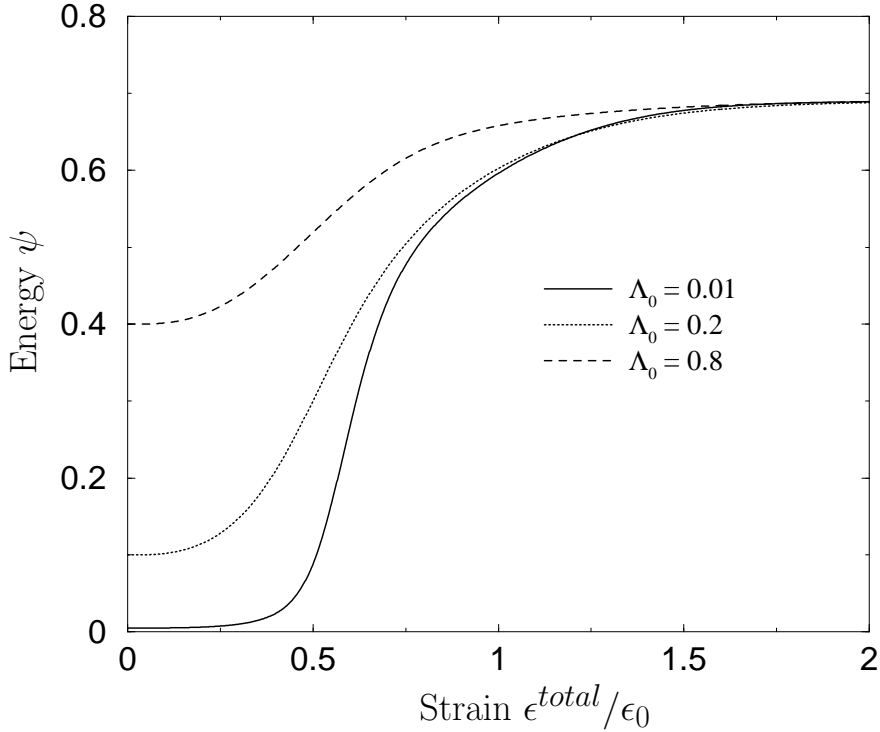}
      \caption{Recoverable energy $\psi$ corresponding to calculations shown 
	in Fig. \ref{fig:stresses}.}
      \label{fig:energy}
\end{figure}

The results shown in Figs. 2 and 3 are qualitatively similar to the experimental data of Hasan and Boyce \cite{HB93,HB95} and of Marano and Rink \cite{MARANO}, both of which groups measured the response of polymeric materials to compressive stress.  Like theirs, our stress-strain curves in Fig. 2 show characteristic peaks and subsequent strain softening. The peak stress is most pronounced for the more highly annealed specimens, which correspond in our language to lower values of $\Lambda_0$, i.e. smaller initial densities of STZ's. The case $\Lambda_0=0.01$ is in effect the limit of perfect annealing.  In contrast, the peak disappears entirely at $\Lambda_0=0.8$. The peak occurs because, when the initial density of STZ's is small, the plastic strain rate must also be small, and the fixed total strain rate must be produced largely by the elastic response to increasing stress. As a result, the stress in the more highly annealed cases shown here initially rises above the yield stress.  Softening then occurs when $\Lambda$ becomes large enough to permit substantial plastic flow.  Note that the nominal STZ yield stress, $s=1$, is not the peak stress but, rather, is the steady-state stress at large strain in the limit of vanishing strain rate. (See Fig. 4 and the discussion below.) 
\begin{figure}
      \includegraphics[angle=-90, width=0.7\columnwidth]{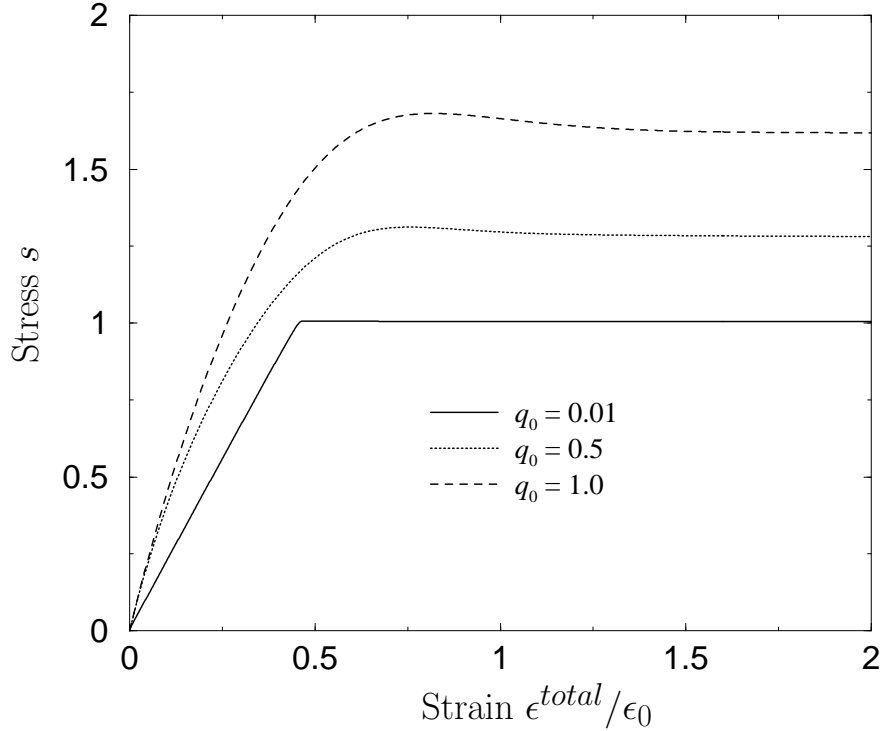}
      \caption{Stress-strain curves for three different strain rates.}
      \label{fig:stress_rate}
\end{figure}

The stored-energy curves shown in Fig. 3 look qualitatively like those shown in \cite{HB93}, where they have been obtained by calorimetric techniques. Measurements of this kind, supplementing the purely mechanical tests, may be especially useful for probing more detailed features of STZ theories.  

In Fig. 4, the case $q_0 = 0.01$ is effectively the limit of zero strain rate.
That stress-strain curve looks like a conventional perfectly elastic - perfectly plastic model; but in fact it is not.  The slope of the ``elastic'' section can be shown to be, not $2\,\mu$, but rather, $2\,\mu/(1+2\,\mu\,\epsilon_0)$.  This is one example of a common feature of the quasilinear STZ theory -- that plastic yielding may occur at all stresses, even those well below the yield stress, depending upon the internal state of the system as characterized by $\Lambda$ and $\Delta$.  Note that the plastic part of this limiting stress-strain curve lies exactly at the yield stress, $s=1$, as expected.  The other two curves in Fig. 4, for larger strain rates, illustrate that this model exhibits a substantial strain-rate sensitivity, perhaps too large a sensitivity as we shall mention below.  

Next consider a series of creep tests in which the strain is measured while the system is loaded to a stress, say, $s_0$ and then held at that stress for an indefinitely long time.  That is, we solve Eq.(\ref{q0}) in the form
\begin{equation}
\dot\epsilon^{total}={\dot s\over 2\,\mu}+ \epsilon_0\,(\Lambda\,s-\Delta),
\end{equation}
where now $s(t)$ is a predetermined function of time $t$.  Specifically, we let $s(t)$ rise linearly from zero to a value $s_0$ in a time interval $\Delta t = 1$. The relevant numerical results are shown in Fig. 5 for the case $\epsilon_0= 0.025$ and $\mu=100$ (consistent with the values chosen above for the constant strain-rate calculations).  
\begin{figure}
      \includegraphics[angle=-90, width=0.7\columnwidth]{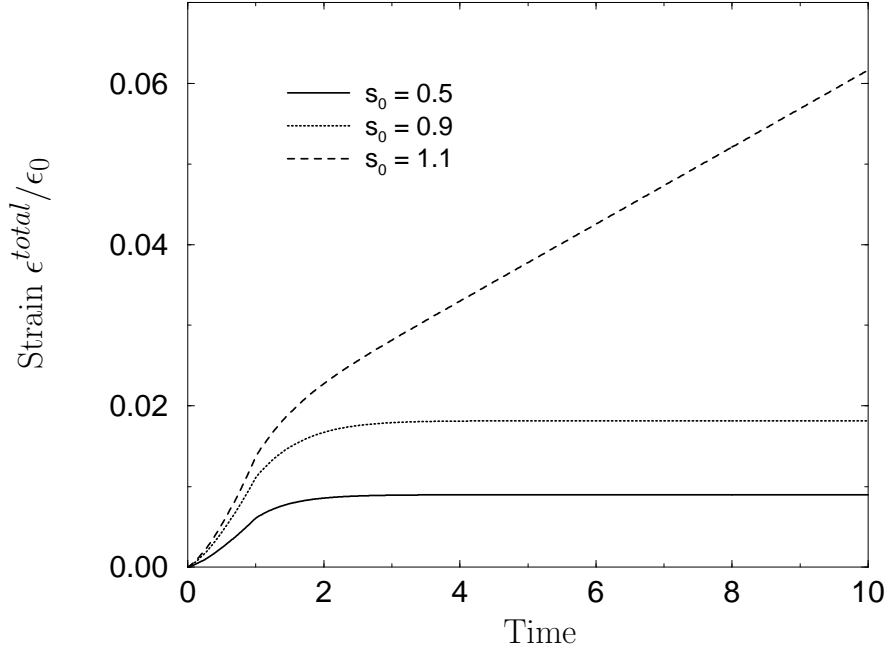}
      \caption{Creep tests for three different final stresses.}
      \label{fig:creep}
\end{figure}
We also choose $\Lambda_0=1$, which maximizes the early plastic response.  Clearly, the system becomes jammed -- the strain rate vanishes -- for stresses $s_0<1$; and, conversely, the strain rate is nonzero for $s_0>1$.  A notable feature here is that, unlike the STZ models discussed in \cite{FL} and \cite{FLMRS}, this version of the quasilinear model does not have a diverging time scale near the yield stress.  We can see this property by noting that the denominator $1-\Delta^2$ on the right-hand side of Eq.(\ref{dotdeltaql2}) did not appear in the earlier theories.  This quantity vanishes as $s\to 1$ along the jammed steady state with $\Delta = s$ or along the flowing state with $\Delta= 1/s$.  Thus, when we linearize this equation about either of those states, the relaxation rate that previously vanished as $s\to 1$ now becomes just unity.  

This version of the STZ theory also exhibits strain recovery on unloading; in fact, the effect is exaggerated. Suppose that we have reduced the stress to zero so rapidly that no plastic response has taken place and $\Delta$ retains the value that it had in the stressed state. Also suppose for simplicity that $\Lambda =1$.  Then the equation of motion for $\Delta$, that is, Eq.(\ref{dotdeltaql2}) for $s=0$, becomes 
\begin{equation}
\dot\Delta = -{\Delta\over(1-\Delta^2)}.
\end{equation}
Clearly, $\Delta$ decreases exponentially to zero on a time scale of order unity. The associated decrease in the plastic strain can be computed from Eq.(\ref{dotepsql}) once we know $\Delta(t)$.  The situation is only slightly more complicated if unloading occurs slowly and some plastic strain recovery takes place before $s$ vanishes.  The important point is that the total strain recovery in this theory depends on the unloading rate.  This history dependence of the recovered plastic strain suggests that it is not an intrinsic anelastic property of the deformed system as suggested in \cite{MARANO}. 

We illustrate these effects in Fig. 6 by showing stress-strain curves for two cases in which the system is first loaded to $s_0$ as in Fig. 5, is later unloaded, and then loaded again.  The specific loading history is shown in the inset.  All parameters are the same as those used in computing Fig. 5. We have chosen the cases $s_0=0.9$ and $s_0=1.1$ for use here in order to compare behaviors of jammed and unjammed systems.  Strain recovery during unloading as well as at $s=0$ is apparent in both cases.
\begin{figure}
      \includegraphics[angle=-90, width=0.7\columnwidth]{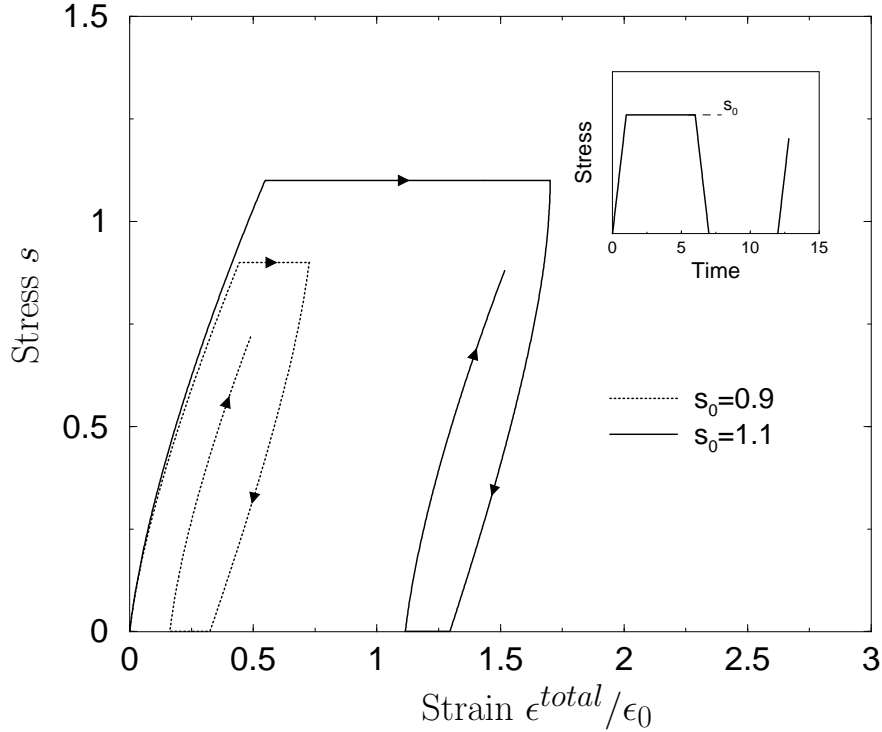}
      \caption{Stress-strain curves for two loading histories.}
      \label{fig:cycling}
\end{figure}
 
The preceding discussion of strain recovery illustrates the loss of memory effects in the quasilinear theory.  Because orientational memory is carried here by the state variable $\Delta$, the fact that $\Delta$ vanishes on a time scale of order unity implies that the system loses memory of its deformed state on the same time scale as that which characterizes plastic response to driving forces.  In a fully nonlinear theory such as that described in \cite{FL}, the transition rates $R_{\pm}(s)$ that determine ${\cal C}(s)$ via Eq.(\ref{Tdef}) may (depending on choice of parameters) become very small when the stress vanishes.  Thus the STZ population after unloading may retain the orientation that it had in its previous stressed state.  This is not a fatal shortcoming; it is possible to fix within the quasilinear framework if desired, but it seems better to use a fully nonlinear theory when the memory effects are of special interest. 

There are other experimental observations that are not accounted for in this minimal version of the STZ theory.  These shortcomings are informative because they point to places where the minimal theory is missing some ingredients.  

One potentially important disagreement is in the predicted steady-state relation between stress and plastic strain rate, which we obtain by setting $\Lambda=1$ and $\Delta = 1/s$ in Eq.(\ref{dotepsql}):
\begin{equation}
\dot\epsilon^{plast} = \epsilon_0\,\cases{0 &for $0<s<1$,\cr (s^2-1)/s &for $s>1$.}
\end{equation} 
This is essentially a Bingham law, that is, the STZ strain rate rises linearly above the yield stress. Many measurements, even in granular materials, indicate a more rapid increase of the form $\dot\epsilon^{plast}\sim s^m$, where $m$ may be large.  Equivalently, stress-strain curves measured at constant strain rate such as those shown in Figs. 2 and 4 often show very little dependence on the strain rate.  

A second interesting discrepancy is that our curves of strain versus time for constant stress (creep tests) shown in Fig. 5 look qualitatively different from those shown by Hassan and Boyce \cite{HB95} in their Figs. 4 and 5.  Specifically, over a range of stresses near the  yield stress, their systems remain jammed at an apparently constant strain for some time, but eventually start to flow plastically.  The delay time for the onset of rapid plastic flow decreases as the stress increases. Such behavior indicates the existence of a new physical mechanism with its own characteristic time scale.  

There are many plausible candidates for additional ingredients or mechanisms that might be added to the minimal STZ model in order to account for these discrepancies. We close this paper by listing some of those that we expect will be important in further investigations.

{\it Shear banding}: In \cite{SHEARLOC}, we pointed out that the STZ theory, when extended to include elastic interactions between the zones, predicts an instability against formation of microstructural shear bands at stresses somewhat lower than the yield stress. The delayed onset of shear banding produced theoretical creep-test results that looked qualitatively like the experiments. In general, shear banding is a phenomenon that will need to be taken into account in interpreting many, if not most, experiments of this kind.  Since publishing \cite{SHEARLOC}, we have found that the STZ theory exhibits shear banding in a wide variety of circumstances.  We hope to report on those investigations in future publications.

The problem of understanding spatial localization of plastic flow in shear banding is closely related to the issue of missing length scales in plasticity theories.  Like almost all other theories of plastic deformation in solids, the version of the STZ theory described here contains no intrinsic length scale.  For example, there are no terms comparable to the viscosity in fluid dynamics or the gradient energy in the Ginzburg-Landau or Cahn-Hilliard theories, both of which determine the scales for spatial variations of the relevant fields.  Without some such term, no plasticity theory can predict an intrinsic width for a shear band or the spatial variation of the shear flow between the interior of the band and the non-deforming material outside of it.  If the STZ picture of molecular rearrangements is realistic, then it ought to help us identify plausible candidates for these length scales.  

{\it Polymer chain dynamics}: All of the experimental data to which we have referred here pertains to amorphous polymeric materials. Equations of motion for polymeric properties such as stretching or entangling will have to be included in any attempt to produce a quantitative description of polymer plasticity. We have not yet included degrees of freedom describing polymeric configurations in any version of the STZ theory; and we expect that we would have to do so in order to achieve quantitative agreement with experiments such as those of \cite {HB93,HB95}.   For example, the present version of the STZ theory could not account for the later-stage strain hardening seen in those experiments.  It would be useful for testing the STZ theory to have comparable experimental data for non-polymeric materials such as metallic glasses.  

{\it Dilation and disorder}: In the original derivation of the STZ theory\cite{FL}, the free volume -- as an intensive variable, conjugate to the true volume and thus roughly analogous to temperature -- played a prominent role in determining transformation rates.  The free volume, however, was treated as a fixed quantity, not as an internal state variable with its own equation of motion.  There are many reasons to believe that shear flow is accompanied by dilation or increased glassy disorder in the form of density fluctuations.  (See, for example, \cite{HUFNAGEL})  Thus, it seems an essential next step in this program to incorporate dynamical measures of dilation or disorder into the STZ analysis.  Lemaitre has proposed one interesting way of doing this.\cite{LEMAITRE}  

{\it Effective temperature}: Finally, we remark that the STZ picture ought to be useful in theories of granular materials, soils, or foams, where the conventional concept of temperature is irrelevant.  There is increasing evidence that the flowing states of such systems are meaningfully characterized by an effective temperature that determines fluctuations and energy flow. \cite{Teff,CUGLIANDOLOetal,SOLLICHetal,BERTHIER-BARRAT}  If that is true, then the effective temperature would also be an important ingredient in theories of plastic flow in conventional molecular materials; it might even be more important for describing the deforming states of these systems than the thermodynamic temperature.  

\begin{acknowledgments}
This research was supported primarily by U.S. Department of Energy Grant
No. DE-FG03-99ER45762. It was also supported in part by the MRSEC Program of the NSF under Award No. DMR96-32716 and by a grant from the Keck Foundation for Interdisciplinary Research in Seismology and Materials Science. 
\end{acknowledgments}

\end{document}